\documentclass[12pt]{article}
\usepackage{amsmath}
\usepackage{amssymb}
\usepackage{amsfonts}
\usepackage{latexsym}
\usepackage{color}
\usepackage{graphicx}

\catcode `\@=11 \@addtoreset{equation}{section}

\catcode `\@=12

\newcommand{\be}{\begin{equation}}
\newcommand{\en}{\end{equation}}
\newcommand{\bea}{\begin{eqnarray}}
\newcommand{\ena}{\end{eqnarray}}
\newcommand{\beano}{\begin{eqnarray*}}
\newcommand{\enano}{\end{eqnarray*}}
\newcommand{\bee}{\begin{enumerate}}
\newcommand{\ene}{\end{enumerate}}
\newcommand{\R}{R \!\!\!\! R}

\newcommand{\Hil}{{\cal H}}

\newcommand{\F}{{\cal F}}
\newcommand{\Lc}{{\cal L}}
\newcommand{\Sc}{{\cal S}}
\newcommand{\D}{{\cal D}}

\newcommand{\G}{{\cal G}}
\newcommand{\E}{{\cal E}}

\newcommand{\1}{1 \!\!\! 1}
\newtheorem{thm}{Theorem}

\newtheorem{prop}[thm]{Proposition}

\newenvironment{proof}{\noindent {\bf Proof:}}{\hfill$\Box$}

\begin{document}

\thispagestyle{empty}

\vspace*{1cm}

\begin{center}
{\Large \bf Gabor-like systems in $\Lc^2(\Bbb{R}^d)$ and extensions to wavelets}   \vspace{2cm}\\

{\large F. Bagarello}\\
  Dipartimento di Metodi e Modelli Matematici,
Fac. Ingegneria, Universit\`a di Palermo, I-90128  Palermo, Italy\\
e-mail: bagarell@unipa.it
\end{center}

%\newpage

\vspace*{2cm}

\begin{abstract}
\noindent In this paper we show how to construct a certain class of
orthonormal bases in $\Lc^2(\Bbb{R}^d)$ starting from one or more
Gabor orthonormal bases in $\Lc^2(\Bbb{R})$. Each such basis can
 be obtained acting on a single function  $\Psi(\underline
x)\in\Lc^2(\Bbb{R}^d)$ with a set of unitary operators which operate
as translation and modulation operators {\em in suitable variables}.
The same procedure is also extended to frames and  wavelets. Many
examples are discussed.

\end{abstract}

\vspace{2cm}


\vfill

\newpage

% Section 1
\section{Introduction}

Gabor and wavelets systems of square-integrable functions are maybe
the most interesting examples of {\em (sometimes)} orthonormal
(o.n.) and complete systems in $\Lc^2(\Bbb{R})$. This is both for
purely mathematical reasons but also in view of their applications
to many different fields of physics, see for instance
\cite{feic,pan}. This produced a deep interest in these subjects and
many mathematicians (but not only) have been involved in their
analysis. In particular, one of the hardest problem to be considered
was the construction of these systems and the analysis of their
properties. This problem has been solved for wavelets by Mallat,
\cite{mall}, introducing the so called multi-resolution analysis.
Since then, many other results have been found and many examples of
o.n. wavelet bases have been constructed. Concerning Gabor systems,
they have received much attention as well and their analysis has
produced results which are often related to coherent states and
frames. Necessary conditions for a Gabor frame to be an o.n. basis
in $\Lc^2(\Bbb{R})$ can be found, for instance, in \cite{dau} and
references therein. A detailed analysis of Gabor frames can be found
in \cite{chr}.

In two recent papers, \cite{abt,bagtri}, the authors  discussed the
possibility of getting an o.n. Gabor basis in $\Lc^2(\Bbb{R})$ (or
in some closed subspace) starting from an non orthogonal Gabor
system. This was shown to be possible under suitable {\em density}
conditions and a rather natural perturbative scheme has been
proposed.

Here we continue our analysis of Gabor systems and we show how a non
trivial Gabor-like o.n. basis in $\Lc^2(\Bbb{R}^d)$ can be
constructed starting from one or more o.n. Gabor bases in
$\Lc^2(\Bbb{R})$. We call this basis {\em Gabor-like} since it can
be still obtained acting on a single function $\Psi(\underline
x)\in\Lc^2(\Bbb{R}^d)$ with a set on unitary operators which behave
like modulation and translation operators but only  {\em in suitable
variables}. We will clarify this point in the following. We also
extend these results to general frames and to wavelets. More in
details, the paper is organized as follows:

In the next section we introduce the (physical and) mathematical
framework which is used in the rest of the paper.

In Section III we show how to construct an o.n. Gabor-like o.n.
basis in two dimensions starting from two Gabor one-dimensional o.n.
bases, which can coincide or not. We further extend this procedure
to  arbitrary dimensions and to frames.

Section IV is devoted to some examples of our construction.

In Section V we adapt the same  strategy to produce o.n.
wavelet-like bases in $\Lc^2(\Bbb{R}^d)$ starting from $d$ o.n.
wavelets bases in $\Lc^2(\Bbb{R})$. As before, we call these new
sets {\em wavelets-like bases} since they are constructed starting
from a single square integrable function $\Psi(\underline x)$ acting
on this with a set on unitary operators which behave like dilation
and translation operators {\em in suitable variables}, as before. We
end the section with some examples of our construction.

Section VI contains our conclusions and projects for the future.

\section{The mathematical setting}

This section is devoted to the construction of the mathematical
setting which is going to be used in our analysis. In what follows
we will only consider a two-dimensional situation. The details of
the extension to higher dimensionality are trivial and are left to
the reader.

Consider the operators $((\hat x_1^{(o)}, \hat p_1^{(o)}),(\hat
x_2^{(o)}, \hat p_2^{(o)}))$ and $((\hat x_1^{(n)}, \hat
p_1^{(n)}),(\hat x_2^{(n)}, \hat p_2^{(n)}))$, acting on a certain
Hilbert space $\Hil$ and satisfying \be[\hat x_j^{(\alpha)},\hat
p_k^{(\alpha)}]=i\,\delta_{jk}, \label{21}\en
where $j,k=1,2$ and $\alpha=o,n$. Here $o$ and $n$ stand for
{\em old} and {\em new}, since what we are interested in is to
consider a unitary transformation from the  canonical operators
$(\hat x_j^{(o)}, \hat p_j^{(o)})$, the {\em old} operators, into
the new ones, $(\hat x_j^{(n)}, \hat p_j^{(n)})$, $j=1,2$.  We can now introduce the {\em generalized
eigenvectors} of all these position operators: \be \hat
x_j^{(\alpha)}\xi_{x,[j]}^{(\alpha)}=x\,\xi_{x,[j]}^{(\alpha)}
\label{22}\en $j=1,2$, $\alpha=o,n$. Then we put, calling
$\underline x=(x_1,x_2)$, \be \xi_{\underline
x}^{(\alpha)}=\xi_{x_1,[1]}^{(\alpha)}\otimes\xi_{x_2,[2]}^{(\alpha)},
\label{23}\en where, as before, $\alpha=o,n$. These vectors satisfy
the following well known properties, \cite{mess}: \be
<\xi_{\underline x}^{(\alpha)},\xi_{\underline
y}^{(\alpha)}>_{\Hil}=\delta(\underline x-\underline
y):=\delta(x_1-y_1)\,\delta(x_2-y_2)\label{24}\en and
\be\int_{\Bbb{R}^2}\,d\underline x\,|\xi_{\underline
x}^{(\alpha)}><\xi_{\underline
x}^{(\alpha)}|=\1_{\Hil},\label{25}\en where we are adopting the
Dirac bra-ket notation. Here $\alpha=o,n$ and we have introduced the
subscript $\Hil$ since in the following we will need to distinguish
between different scalar products in different Hilbert spaces. Any
element $\Psi\in\Hil$ can be represented using the {\em new} or the
{\em old} eigenvectors: \be \Psi^{(\alpha)}(\underline
x):=<\xi_{\underline x}^{(\alpha)},\Psi>_{\Hil}, \qquad
\alpha=o,n\label{26}\en This is a well-known procedure in quantum
mechanics which corresponds to the possibility of using different
representations (like the position and the momentum representations)
to describe the same physical vector. It is quite easy to prove
using (\ref{25}) that,  even if the generalized eigenvectors introduced in (\ref{22})-(\ref{23}) do not belongs to $\Hil$,  both $\Psi^{(o)}(\underline x)$ and
$\Psi^{(n)}(\underline x)$ belong to $\Lc^2(\Bbb{R}^2)$ and that
$\|\Psi^{(n)}(\underline
x)\|_{\Lc^2(\Bbb{R}^2)}=\|\Psi^{(o)}(\underline
x)\|_{\Lc^2(\Bbb{R}^2)}=\|\Psi\|_{\Hil}$ for all $\Psi\in\Hil$. The two functions
$\Psi^{(n)}(\underline x)$ and $\Psi^{(o)}(\underline x)$ are
related to each other via the following equations:
\be\Psi^{(o)}(\underline x)=\int\,d\underline y\,<\xi_{\underline
x}^{(o)},\,\xi_{\underline y}^{(n)}>_{\Hil}\Psi^{(n)}(\underline
y)=\int\,d\underline y\,K(\underline x;\underline
y)\,\Psi^{(n)}(\underline y)\label{27}\en and
\be\Psi^{(n)}(\underline x)=\int\,d\underline y\,<\xi_{\underline
x}^{(n)},\,\xi_{\underline y}^{(o)}>_{\Hil}\Psi^{(o)}(\underline
y)=\int\,d\underline y\,\overline{K(\underline y;\underline
x)}\,\Psi^{(o)}(\underline y),\label{28}\en where we have introduced
the {\em kernel} of the transformation, \be K(\underline
x;\underline y)=<\xi_{\underline x}^{(o)},\,\xi_{\underline
y}^{(n)}>_{\Hil}\label{29}\en

Since the operators $\hat x_j^{(\alpha)}$ and $\hat p_k^{(\beta)}$ are unbounded, they
 act on a dense domain $\D_\Hil$ of $\Hil$. Moreover, the action of, say, $\hat x_j^{(\alpha)}$ on the set $\D_{gen}$ of the generalized eigenvectors introduced in (\ref{22})-(\ref{23}) is also well defined, see (\ref{22}) and \cite{mess}. For this reason the matrix element $<f,\hat x_j^{(\alpha)}g>_{\Hil}$, which makes no sense
for general $f,g\in\Hil$,  is well defined if, for instance, $g$  belongs to $\D_{gen}$ or to $\D_\Hil$ and $f\in\Hil$. In the first case, i.e. if we consider $<f,\hat x_j^{(\alpha)}\xi_{\underline
x}^{(\alpha)}>_{\Hil}$, the result of this computation is a function of $\underline x$, $\,x_j<f,\xi_{\underline
x}^{(\alpha)}>_{\Hil}=x_j\,\overline{f^{(\alpha)}(\underline
x)}$, which is not necessarily square integrable (while $f^{(\alpha)}(\underline
x)\in\Lc^2(\Bbb{R}^2)$). We can now introduce other
 operators, $\hat X_j^{(\alpha)}$ and $\hat P_k^{(\alpha)}$, which we
often call in the rest of the paper the {\em capital operators}, via
the  equations:
 \be
\hat X_j^{(\alpha)}<f,g>_{\Hil}:=<f,\hat x_j^{(\alpha)}g>_{\Hil},
\qquad \hat P_j^{(\alpha)}<f,g>_{\Hil}:=<f,\hat
p_j^{(\alpha)}g>_{\Hil},\label{210}\en for all $\alpha$ and $j$, where $f$ and
$g$ are as above, and in particular they are such that the right-hand sides in (\ref{210}) are well defined. For instance we have $\hat X_j^{(\alpha)}<f,\xi_{\underline
x}^{(\alpha)}>_{\Hil}=<f,\hat x_j^{(\alpha)}\xi_{\underline
x}^{(\alpha)}>_{\Hil}=x_j<f,\xi_{\underline
x}^{(\alpha)}>_{\Hil}$.
 These new operators are still
canonically conjugate since they satisfy $[\hat X_j^{(\alpha)},\hat
P_k^{(\alpha)}]=i\,\delta_{jk}$ (on a suitable
domain).

The definition above produces a particularly relevant expression if
we introduce the following unitary operators: \be t_+(\underline
a)=e^{i\hat{\underline x}^{(n)}\cdot\underline a},\qquad
t_-(\underline b)=e^{i\hat{\underline p}^{(n)}\cdot\underline
b}\label{211}\en and their capital counterparts $ T_+(\underline
a)=e^{i\hat{\underline X}^{(n)}\cdot\underline a}$ and
$T_-(\underline b)=e^{i\hat{\underline P}^{(n)}\cdot\underline b}$.
Here $\underline a=(a_1,a_2)$, $\underline b=(b_1,b_2)$,
$\hat{\underline X}^{(n)}=(\hat X_1^{(n)},\hat X_2^{(n)})$, $\hat{\underline P}^{(n)}=(\hat P_1^{(n)},\hat P_2^{(n)})$, $\hat{\underline x}^{(n)}\cdot\underline a=\hat x_1^{(n)}a_1+\hat x_2^{(n)}a_2$, and so
on. In this case we have, for instance, \be T_+(\underline
a)f^{(\alpha)}(\underline x)= T_+(\underline a)<\xi_{\underline
x}^{(\alpha)},\,f>_\Hil=<\xi_{\underline
x}^{(\alpha)},t_+(\underline a)\,f>_\Hil \label{212}\en and \be
T_-(\underline b)f^{(\alpha)}(\underline x)= T_-(\underline
b)<\xi_{\underline x}^{(\alpha)},\,f>_\Hil=<\xi_{\underline
x}^{(\alpha)},t_-(\underline b)\,f>_\Hil, \label{213}\en for all
$f\in\Hil$, $\alpha=o,n$.

As it is clear the operators $T_+(\underline a)$ and $T_-(\underline
b)$ are modulation and translation  acting on $\Lc^2(\Bbb{R}^2)$
(but in the new variables, see below). Moreover they satisfy \be
T_+(\underline a)T_-(\underline b)=e^{-i\underline a\cdot\underline
b}\,T_-(\underline b)T_+(\underline a),\label{214}\en and an
analogous equality holds for $t_+(\underline a)$ and $t_-(\underline
b)$.

\vspace{2mm}

{\bf Remark:} it is worth noticing that we are now introducing an
asymmetry between the {\em old} and the {\em new} operators since,
for reasons that will appear clearly in the next section, we just
need to define those unitary operators  associated to $(\hat
x_j^{(n)}, \hat p_j^{(n)})$ and $(\hat X_j^{(n)}, \hat P_j^{(n)})$,
and not to the old ones.

\section{Gabor-like o.n. bases in any dimension}

The question we now want to address is quite a natural one and looks
as follows: calling $\underline a(\underline l)=(a_1l_1,a_2l_2)$ and
$\underline b(\underline k)=(b_1k_1,b_2k_2)$ we ask whether it is
possible to find a square-integrable function, which we call
$\Psi^{(o)}(\underline x)\in\Lc^2(\Bbb{R}^2)$ for reasons that will
appear clear in a moment, such that the set \be
\Sc=\left\{\Psi^{(o)}_{\underline l,\underline k}(\underline
x)=T_+(\underline a(\underline l))\,T_-(\underline b(\underline
k))\,\Psi^{(o)}(\underline x),\,\underline k,\underline
l\in\Bbb{Z}^2\right\} \label{31}\en is an o.n. basis in
$\Lc^2(\Bbb{R}^2)$.

We like to stress that the problem is only close but not identical
to the construction of a Gabor o.n. basis in $\Lc^2(\Bbb{R}^2)$. This is because, as we have already remarked before, the
$T_{\pm}$ operators depends on the {\bf new} operators while
$\Psi^{(o)}(\underline x)=<\xi_{\underline x}^{(o)},\Psi>_{\Hil}$ is
the projection of $\Psi$ on the generalized eigenvectors of the {\bf old}
position operators. For this reason the set $\Sc$ is not obtained
from $\Psi^{(o)}(\underline x)$  via {\em canonical} translations
and modulations and this is why we call $\Sc$ a {\em Gabor-like}
rather than simply a {\em Gabor} set. This will be made explicit in
the next section, when many examples will be discussed.

As already mentioned this problem is quite close to that addressed
in \cite{bagtri,abt}, where the authors have discussed a general strategy to
construct, starting from a non o.n. set
$\{f_{k_1,\ldots,k_N}:=A_1^{k_1}\cdots A_N^{k_N}f_0,\,\, k_1,\ldots
k_N\in\Bbb{Z} \}$ where $f_0\in\Hil$ is a fixed element and
$A_1,\ldots,A_N$  are $N$ given unitary operators, a different o.n.
set, again of the same form $\{A_1^{k_1}\cdots A_N^{k_N}h_0,\,\,
k_1,\ldots k_N\in\Bbb{Z} \}$. For that we have discussed how to get
such a vector $h_0\in\Hil$ and we have shown that the procedure of
orthogonalization usually fails to work in all of $\Hil$ while it
works perfectly in suitable closed subspaces of $\Hil$, where the
perturbation expansions adopted in \cite{bagtri,abt} converge quite
fast. In the present settings the unitary operators are  $T_+$ and
$T_-$ and our unknown is the function $\Psi^{(o)}(\underline x)$.
The procedure we are going to develop here is rather general,
non-perturbative, and produces o.n. bases for all of
$\Lc^2(\Bbb{R}^2)$.

\vspace{2mm}

Let us  now go back to our original question. First we rewrite
$\Psi^{(o)}_{\underline l,\underline k}(\underline x)$ in a
convenient form. This can be done as follows:
$$
\Psi^{(o)}_{\underline l,\underline k}(\underline x)=T_+(\underline
a(\underline l))\,T_-(\underline b(\underline
k))\,\Psi^{(o)}(\underline x)=<\xi_{\underline
x}^{(o)},t_+(\underline a(\underline l))\,t_-(\underline
b(\underline k))\,\Psi>_\Hil=
$$
$$
=\int_{\Bbb{R}^2}\,d\underline y <\xi_{\underline
x}^{(o)},\,\xi_{\underline y}^{(n)}>_\Hil<\xi_{\underline
y}^{(n)},t_+(\underline a(\underline l))\,t_-(\underline
b(\underline k))\,\Psi>_\Hil=$$ \be\int_{\Bbb{R}^2}\,d\underline y\,
K(\underline x;\underline y)\,e^{i\underline y\cdot \underline
a(\underline l)}\Psi^{(n)}(\underline y+ \underline b(\underline k))
\label{32}\en where we have used (\ref{29}), (\ref{212}),
(\ref{213}), the resolution of the identity for $\{\xi_{\underline
y}^{(n)}\}$, and the fact that, as it easily checked,
$t_-(\underline b(\underline k))^\dagger\,t_+(\underline
a(\underline l))^\dagger\,\xi_{\underline y}^{(n)}=e^{-i\underline
y\cdot \underline a(\underline
l)}\,\xi_{\left(y_1+b_1k_1,y_2+b_2k_2\right)}^{(n)}=e^{-i\underline
y\cdot \underline a(\underline l)}\,\xi_{\underline y+ \underline
b(\underline k)}^{(n)}$.

The next step consists now in computing the following scalar product
in $\Lc^2(\Bbb{R}^2)$: $$\Omega_{\underline l,\,\underline
k}:=<\Psi^{(o)}_{\underline l,\underline
k},\Psi^{(o)}>_{\Lc^2(\Bbb{R}^2)}=\int_{\Bbb{R}^2}\,d\underline
x\,\overline{\Psi^{(o)}_{\underline l,\underline k}(\underline
x)}\,\Psi^{(o)}(\underline x)$$ Using formula (\ref{32}) and the
resolution of the identity for the set $\{\xi_{\underline
x}^{(o)}\}$ we deduce that \be \Omega_{\underline l,\,\underline
k}=\int_{\Bbb{R}^2}\,d\underline y\,e^{-i\underline y\cdot
\underline a(\underline l)}\overline{\Psi^{(n)}(\underline
y+\underline b(\underline k))}\,\Psi^{(n)}(\underline y)
\label{33}\en This last integral can be written in a factorized form
if $\Psi^{(n)}(\underline y)$ is, by itself, factorizable. This does
not necessarily imply that also $\Psi^{(o)}(\underline x)$ can be
written as the product of a function which only depends on $x_1$
times another function which depends on $x_2$. We will come back to
this point in the next section where our examples will show
explicitly this claim. This implies, of course, that our approach is
rather different (and more interesting) than simply taking tensor
products. Let us therefore assume that \be \Psi^{(n)}(\underline
y)=h_1(y_1)\,h_2(y_2)\label{34}\en Hence we find  \be
\Omega_{\underline l,\,\underline
k}=\left(\int_{\Bbb{R}}dy_1\,e^{-iy_1a_1l_1}\,\overline{h_1(y_1+b_1k_1)}\,h_1(y_1)\right)
\left(\int_{\Bbb{R}}dy_2\,e^{-iy_2a_2l_2}\,\overline{h_2(y_2+b_2k_2)}\,h_2(y_2)\right)
\label{35}\en This is the first ingredient for the proof of the
Proposition 1 below. Before stating the main result of this section,
however, we still need to compute the scalar product (in
$\Lc^2(\Bbb{R}^2)$) between a generic element
$\Psi^{(o)}_{\underline l,\underline k}(\underline x)$ of the set
$\Sc$ and a function $f^{(o)}(\underline x)$ of $\Lc^2(\Bbb{R}^2)$.
This computation is not very different from the one in (\ref{33}).
Indeed, the same steps can be repeated and we find that \be
<\Psi^{(o)}_{\underline l,\underline
k},\,f^{(o)}>_{\Lc^2(\Bbb{R}^2)}=\int_{\Bbb{R}^2}\,d\underline
x\,\overline{\Psi^{(o)}_{\underline l,\underline k}(\underline
x)}\,f^{(o)}(\underline x)=\int_{\Bbb{R}^2}\,d\underline
y\,e^{-i\underline y\cdot \underline a(\underline
l)}\overline{\Psi^{(n)}(\underline y+\underline b(\underline
k))}\,f^{(n)}(\underline y) \label{36}\en which gives the scalar
product of the two functions in {\em the new representation}. It is
further convenient to write this equation in a different way, by
introducing the function \be
F_{l_2,k_2}(y_1):=\int_{\Bbb{R}}dy_2\,e^{-iy_2a_2l_2}
\,\overline{h_2(y_2+b_2k_2)}\,f^{(n)}(y_1,y_2)\label{37}\en We get
\be <\Psi^{(o)}_{\underline l,\underline
k},\,f^{(o)}>_{\Lc^2(\Bbb{R}^2)}=\int_{\Bbb{R}}dy_1\,e^{-iy_1a_1l_1}\,
\overline{h_1(y_1+b_1k_1)}\,F_{l_2,k_2}(y_1)\label{38}\en It is
evident from the previous formulas that a relevant role in all our
computations is played by the one-dimensional Gabor systems
generated by $h_1$ and $h_2$:
$s_j=\{\,h^{(j)}_{l,k}(x):=e^{ixa_jl}\,h_j(x+b_jk),\,k,l\in\Bbb{Z}\}$,
$j=1,2$.

The following proposition holds true:

\begin{prop}
Let $s_1$ and $s_2$ be two o.n. bases in $\Lc^2(\Bbb{R})$. Then the
set $\Sc$ is an o.n. basis in $\Lc^2(\Bbb{R}^2)$.
\end{prop}
\vspace{2mm}

\begin{proof}
We first prove that the set $\Sc$ is orthonormal. For that we start
remarking that our assumptions on $s_1$ and $s_2$, together with
(\ref{35}), imply that $\Omega_{\underline l,\,\underline
k}=<\Psi^{(o)}_{\underline l,\underline
k},\Psi^{(o)}>_{\Lc^2(\Bbb{R}^2)}=\delta_{\underline l,\,\underline
0}\,\delta_{\underline k,\,\underline 0}$. This result, together
with easy properties of the operators $T_{\pm}$, implies in turns
$<\Psi^{(o)}_{\underline l,\underline k},\Psi^{(o)}_{\underline
n,\underline m}>_{\Lc^2(\Bbb{R}^2)}=\delta_{\underline
l,\,\underline n}\,\delta_{\underline k,\,\underline m}$.

To prove that $\Sc$ is complete in $\Lc^2(\Bbb{R}^2)$ we assume that
$<\Psi^{(o)}_{\underline l,\underline
k},\,f^{(o)}>_{\Lc^2(\Bbb{R}^2)}=0$ for all $\underline l,\underline
k\in\Bbb{Z}^2$. Hence, because of (\ref{38}) and of the completeness
of the set $s_1$, $F_{l_2,k_2}(y_1)=0$ almost everywhere (a.e.) in
$\Bbb{R}$ for all possible choices of $l_2$ and $k_2$ in $\Bbb{Z}$.
As a consequence of this,  of (\ref{37}) and of the completeness of
$s_2$, we conclude that also the function
$g_{y_1}(y_2):=f^{(n)}(y_1,y_2)$ is zero a.e., which finally implies
that $f^{(o)}(\underline x)=\int_{\Bbb{R}^2}\,d\underline
y\,K(\underline x;\underline y)\,f^{(n)}(\underline y)$ is zero a.e.
as well.

\end{proof}

\vspace{2mm}

{\bf Remark:} it might be worth remarking once again that the
procedure is only apparently trivial since, even if
$\Psi^{(n)}(\underline y)$ is factorized, $\Psi^{(o)}(\underline x)$
needs not to be. We will show this in the examples discussed in the
next section.

\vspace{2mm}

\subsection{Extensions to $d>2$ and to frames}

It is not very hard to imagine that this procedure can be extended
to any dimension. In this case we start with a set of $d$ {\em old}
conjugate operators $(\hat x_j^{(o)}, \hat p_j^{(o)})$,
$j=1,2,\ldots,d$, which are mapped, via a canonical transformation,
into a {\em new} set of $d$ conjugate operators $(\hat x_j^{(n)},
\hat p_j^{(n)})$, $j=1,2,\ldots,d$. As before we can introduce the
generalized eigenvectors of all these position operators, their
tensor product $\xi_{\underline
x}^{(\alpha)}=\xi_{x_1,[1]}^{(\alpha)}\otimes\xi_{x_2,[2]}^{(\alpha)}\otimes
\cdots\otimes\xi_{x_d,[d]}^{(\alpha)}$, $\alpha=o,n$, and these vectors
satisfy a $\delta-$like normalization and a resolution of the
identity, see (\ref{24}) and (\ref{25}). Then, if we put
$\Psi^{(\alpha)}(\underline x):=<\xi_{\underline
x}^{(\alpha)},\Psi>_{\Hil}$, we can change between {\em old} and
{\em new} variables and the relations are given in
(\ref{27})-(\ref{29}). Moreover, we can introduce {\em capital} and
{\em small} unitary operators like in (\ref{211})-(\ref{213}). For
instance we have $t_+(\underline a)=e^{i\hat{\underline
x}^{(n)}\cdot\underline a}$, where $\hat{\underline
x}^{(n)}\cdot\underline a=\hat x_1^{(n)}a_1+\cdots+\hat
x_d^{(n)}a_d$.

As before the question is the following: is it possible to find a
function $\Psi^{(o)}(\underline x)\in\Lc^2(\Bbb{R}^d)$ such that the
set $\Sc=\left\{\Psi^{(o)}_{\underline l,\underline k}(\underline
x)=T_+(\underline a(\underline l))\,T_-(\underline b(\underline
k))\,\Psi^{(o)}(\underline x),\,\underline k,\underline
l\in\Bbb{Z}^2\right\}$ is an o.n. basis in $\Lc^2(\Bbb{R}^d)$? Here,
extending our previous notation, we have $\underline a(\underline
l)=(a_1l_1,\ldots,a_dl_d)$ and $\underline b(\underline
k)=(b_1k_1,\ldots,b_dk_d)$. Once again the answer is positive and
the procedure is an almost trivial extension of the one discussed
above. We deduce that
$$
\Psi^{(o)}_{\underline l,\underline k}(\underline
x)=\int_{\Bbb{R}^d}\,d\underline y\, K(\underline x;\underline
y)\,e^{i\underline y\cdot \underline a(\underline
l)}\Psi^{(n)}(\underline y+ \underline b(\underline k)),
$$
where $K(\underline x;\underline y)=<\xi_{\underline
x}^{(o)},\xi_{\underline y}^{(n)}>_{\Hil}$, and, if
$\Psi^{(n)}(\underline y)=h_1(y_1)\cdots h_d(y_d)$, then
$$\Omega_{\underline l,\,\underline k}:=<\Psi^{(o)}_{\underline
l,\underline k},\Psi^{(o)}>_{\Lc^2(\Bbb{R}^d)}=\prod_{j=1}^d\left(
\int_{\Bbb{R}}dy_j\,e^{-iy_ja_jl_j}\,\overline{h_j(y_j+b_jk_j)}\,h_j(y_j)\right),$$
while formula (\ref{38}) must be replaced by \be
<\Psi^{(o)}_{\underline l,\underline
k},\,f^{(o)}>_{\Lc^2(\Bbb{R}^d)}=\int_{\Bbb{R}}dy_1\,e^{-iy_1a_1l_1}\,
\overline{h_1(y_1+b_1k_1)}\,F_{l_2,k_2,\cdots,l_d,k_d}^{(1)}(y_1)\label{39}\en
where \be
F_{l_2,k_2,\cdots,l_d,k_d}^{(1)}(y_1):=\int_{\Bbb{R}^{d-1}}dy_2\ldots
dy_d\,e^{-i(y_2a_2l_2+\cdots+y_da_dl_d)}\times$$
$$\times\overline{h_2(y_2+b_2k_2)}\cdots\overline{h_d(y_d+b_dk_d)}\,
f^{(n)}(y_1,y_2,\ldots,y_d)\label{310}\en Needless to say, in this
case we need to introduce $d$ different one-dimensional Gabor
systems $s_j=\{\,e^{ixa_jl}\,h_j(x+b_jk),\,k,l\in\Bbb{Z}\}$,
$j=1,2,\ldots,d$, and if each one of these is an o.n. basis in
$\Lc^2(\Bbb{R})$\footnote{Of course they could all be coincident. In
this case it is enough to start with a single o.n. Gabor basis in
$\Lc^2(\Bbb{R})$.}, then it is trivial to check that $\Sc$ is an
o.n. set in $\Lc^2(\Bbb{R}^d)$. As for its completeness, our
previous proof can also be extended in this case: from (\ref{39})
and from the completeness of $s_1$ we first deduce that
$F_{l_2,k_2,\cdots,l_d,k_d}^{(1)}(y_1)$ is zero a.e. for all
possible integer choices of $l_2, k_2,\ldots, l_d$ and $k_d$. But,
because of (\ref{310}) and of the completeness of $s_2$, also the
function $$
F_{l_3,k_3,\cdots,l_d,k_d}^{(2)}(y_1,y_2):=\int_{\Bbb{R}^{d-2}}dy_3\ldots
dy_d\,e^{-i(y_3a_3l_3+\cdots+y_da_dl_d)}\times$$
$$\times\overline{h_3(y_3+b_3k_3)}\cdots\overline{h_d(y_d+b_dk_d)}\,
f^{(n)}(y_1,y_2,\ldots,y_d)$$ is zero a.e. for all possible $l_3,
k_3,\ldots, l_d,k_d\in\Bbb{Z}$. Iterating this procedure we finally
conclude that $f^{(n)}(\underline y)=0$ a.e. and therefore
$f^{(o)}(\underline x)=0$.

\vspace{3mm}

We postpone to the next section one example of this construction.
Now we want to show how this same procedure can also be adapted to
construct frames in $d$-dimensions starting from $d$ frames in one
dimension. For reader's convenience we recall that an $(A,B)$-frame of an Hilbert space $\Hil$, $0<A\leq B<\infty$, is a set of vectors $\{\varphi_n\in\Hil, \,n\in J\}$, labeled by a set $J\subseteq\Bbb{N}$,   such that the inequality $ A\|f\|^2 \leq
\sum_{n\in J}|<\varphi_n,f>|^2 \leq B\|f\|^2 $ holds for any
$f\in \Hil$.

In particular we can prove the following

\begin{prop}
Let $s_j$ be an $(A_j,B_j)-$frame, $j=1,2,\ldots,d$, in
$\Lc^2(\Bbb{R})$. Then the set $\Sc$ is an $(A_1A_2\cdots
A_d,B_1B_2\cdots B_d)-$frame in $\Lc^2(\Bbb{R}^d)$.
\end{prop}

\begin{proof}
We give the proof of this statement only for $d=2$. The extension to
higher dimensions is straightforward.

Using (\ref{38}) and the definition of $s_1$ we deduce that
$$
\sum_{\underline k,\,\underline
l\in\Bbb{Z}^2}\,\left|<\Psi^{(o)}_{\underline l,\underline
k},\,f^{(o)}>_{\Lc^2(\Bbb{R}^2)}\right|^2=\sum_{ k_2,\,
l_2\in\Bbb{Z}}\,\left(\sum_{ k_1,\,
l_1\in\Bbb{Z}}\,\left|<h^{(1)}_{l_1,k_1},F_{l_2,k_2}>_{\Lc^2(\Bbb{R})}
\right|^2\right),
$$
so that, because of our assumption on $s_1$, we get
$$
A_1\sum_{ k_2,\,
l_2\in\Bbb{Z}}\,\|F_{l_2,k_2}\|^2_{\Lc^2(\Bbb{R})}\leq
\sum_{\underline k,\,\underline
l\in\Bbb{Z}^2}\,\left|<\Psi^{(o)}_{\underline l,\underline
k},\,f^{(o)}>_{\Lc^2(\Bbb{R}^2)}\right|^2\leq B_1\sum_{ k_2,\,
l_2\in\Bbb{Z}}\,\|F_{l_2,k_2}\|^2_{\Lc^2(\Bbb{R})}
$$
Now, since $s_2$ is an $(A_2,B_2)$-frame, because of (\ref{37}) it
is not hard to prove that
$$
A_2\|f^{(n)}\|^2_{\Lc^2(\Bbb{R}^2)}\leq \sum_{ k_2,\,
l_2\in\Bbb{Z}}\,\|F_{l_2,k_2}\|^2_{\Lc^2(\Bbb{R})}\leq
B_2\|f^{(n)}\|^2_{\Lc^2(\Bbb{R}^2)}
$$
which, together with our previous estimate and using the equality
$\|f^{(n)}\|_{\Lc^2(\Bbb{R}^2)}=\|f^{(o)}\|_{\Lc^2(\Bbb{R}^2)}$,
implies the statement.

\end{proof}

\section{Examples}

This section is devoted to discuss in details some examples,
concerning the construction above of an o.n. Gabor-like basis and of
Gabor-like frame of $\Lc^2(\Bbb{R}^d)$. As we have discussed in the
previous section, and it is clarified by formula (\ref{27}),
$\Psi^{(o)}(\underline x)=\int\,d\underline y\,K(\underline
x;\underline y)\,\Psi^{(n)}(\underline y)$, a special role in our
construction is played by the kernel of the canonical transformation
between $(\hat x_j^{(o)},\hat p_j^{(o)})$ and $(\hat x_j^{(n)},\hat
p_j^{(n)})$, $j=1,\ldots,d$. Of course the analytic expression of
$K(\underline x;\underline y)=<\xi_{\underline
x}^{(o)},\,\xi_{\underline y}^{(n)}>_{\Hil}$ strongly depends on the
transformation itself and it can be deduced using the results in
\cite{mo}. Some of the kernels we are going to consider here are
related to physically relevant transformations, already considered
by the author in different contexts. As for the o.n. Gabor systems
in $d=1$, we will consider here the following two well known sets,
\cite{dau}: \be \G^{(j)}=\left\{g^{(j)}_{m,n}(x)=e^{2\pi i m
x}\,g^{(j)}(x-n), \,m,\,n\in\Bbb{Z}\right\}, \label{40}\en where
$g^{(1)}(x)=\chi_{[0,1]}(x)$ or $g^{(2)}(x)=\frac{\sin(\pi x)}{\pi
x}$. Here $\chi_{[0,1]}(x)$ is the characteristic function of the
interval $[0,1]$.

\subsection{Quantum Hall effect}

In the analysis of the quantum Hall effect a particularly relevant
operator is the hamiltonian of a single electron subjected to a
strong and uniform magnetic field along $z$. This hamiltonian, in
the so-called symmetric gauge, looks like
$$
H=\frac{1}{2}(\underline p+\underline A(\underline
r))^2=\frac{1}{2}\left(p_x-\frac{y}{2}\right)^2+\left(p_y+\frac{x}{2}\right)^2,
$$
which can be written, identifying $x,\,y,\,p_x$ and $p_y$
respectively with $\hat x_1^{(o)},\,\hat x_2^{(o)},\,\hat p_1^{(o)}$
and $\hat p_2^{(o)}$, as $ H=\frac{1}{2}(\hat p_1^{(o)}-\hat
x_2^{(o)}/2)^2+(\hat p_2^{(o)}+\hat x_1^{(o)}/2)^2$. If we now
introduce the following canonical transformation, see \cite{bms} and
references therein, \be \hat x_1^{(n)}=\hat p_1^{(o)}+\frac{\hat
x_2^{(o)}}{2},\quad \hat x_2^{(n)}=\hat p_2^{(o)}+\frac{\hat
x_1^{(o)}}{2},\quad \hat p_1^{(n)}=\hat p_2^{(o)}-\frac{\hat
x_1^{(o)}}{2},\quad \hat p_2^{(n)}=\hat p_1^{(o)}-\frac{\hat
x_2^{(o)}}{2},\quad \label{41}\en $H$ can be written as
$H=\frac{1}{2}((\hat p_2^{(n)})^2+(\hat x_2^{(n)})^2)$. This,
because of the canonicity of the transformation, is just the
hamiltonian of a quantum harmonic oscillator in the variables $(\hat
x_2^{(n)},\hat p_2^{(n)})$. The physical reason for introducing
(\ref{41}) is that it displays rather clearly the infinite
degeneracy of the so-called {\em Landau levels}. This means that, if
$\varphi(x)$ is an eigenstate of $H$ with eigenvalue $E$, then
$e^{i\hat x_1^{(n)}\alpha}e^{i\hat p_1^{(n)}\beta}\varphi(x)$ is
still an eigenstate of $H$, corresponding to the same eigenvalue,
for all possible choices of $\alpha$ and $\beta$. This and many
other aspects of the transformation in (\ref{41}), together with its
possible application to the theory of multi-resolution analysis,
have been discussed in \cite{bag2005} and references therein. The
kernel $K(\underline x;\underline y)$ can be found using the results
in \cite{mo} and in \cite{daza}. We get \be K(\underline
x;\underline
y)=\frac{1}{2\pi}\,\exp\left\{i\left(x_1y_1+x_2y_2-y_1y_2-\frac{x_1x_2}{2}\right)\right\}\label{42}\en
We are now ready to find the function $\Psi^{(o)}(\underline x)$ for
different choices of the {\em one-dimensional ingredients}, see
(\ref{40}), and considering the  formula $ \Psi^{(o)}(\underline
x)=\int\,d\underline y\,K(\underline x;\underline
y)\,\Psi^{(n)}(\underline y)$, with $\Psi^{(n)}(\underline
y)=h_1(y_1)h_2(y_2)$.

We start considering here the following choice:
$h_1(x)=h_2(x)=\chi_{[0,1]}(x)$. This is a symmetric choice and
produces the following function:
$$
\Psi^{(o)}(\underline x)=\frac{e^{ix_1x_2/2}}{2\pi
i}\int_{x_2-1}^{x_2}dt \,e^{-ix_1t}\,\frac{e^{it}-1}{t},
$$
which can be computed explicitly and gives a rather involved
analytic expression for $\Psi^{(o)}(\underline x)$ which {\bf is not
factorizable}. Rather than giving this expression, we just give here
the plot of the modulus of $\Psi^{(o)}(\underline x)$:

\begin{center}
\mbox{\includegraphics[height=6.4cm, width=9.0cm]
{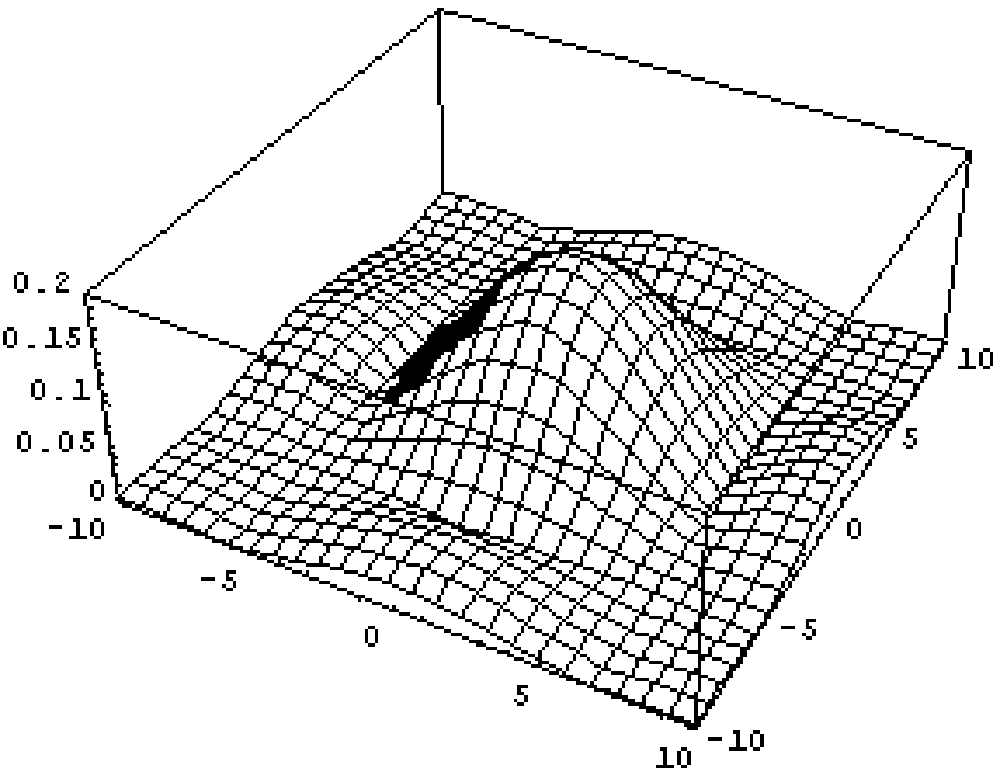}}\hspace{6mm}
\hfill\\
\begin{figure}[h]
\caption{\label{fig1} $|\Psi^{(o)}(\underline x)|$ for
$h_1(x)=h_2(x)=\chi_{[0,1]}(x)$}
\end{figure}
\end{center}

As we can see this function is localized around the origin and goes
to zero when $x^2+y^2$ diverges.

Another example can be constructed choosing
$h_1(x)=h_2(x)=\frac{\sin(\pi x)}{\pi x}$. In this case we get
$$
\Psi^{(o)}(\underline
x)=\frac{e^{-ix_1x_2/2}}{2\pi^2}\int_{x_2-\pi}^{x_2+\pi}dt
\,e^{ix_1t}\frac{\sin(\pi t)}{t},
$$
whose modulus is plotted in the next figure.

\vspace{3mm}

\begin{center}
\mbox{\includegraphics[height=6.4cm, width=9.0cm]
{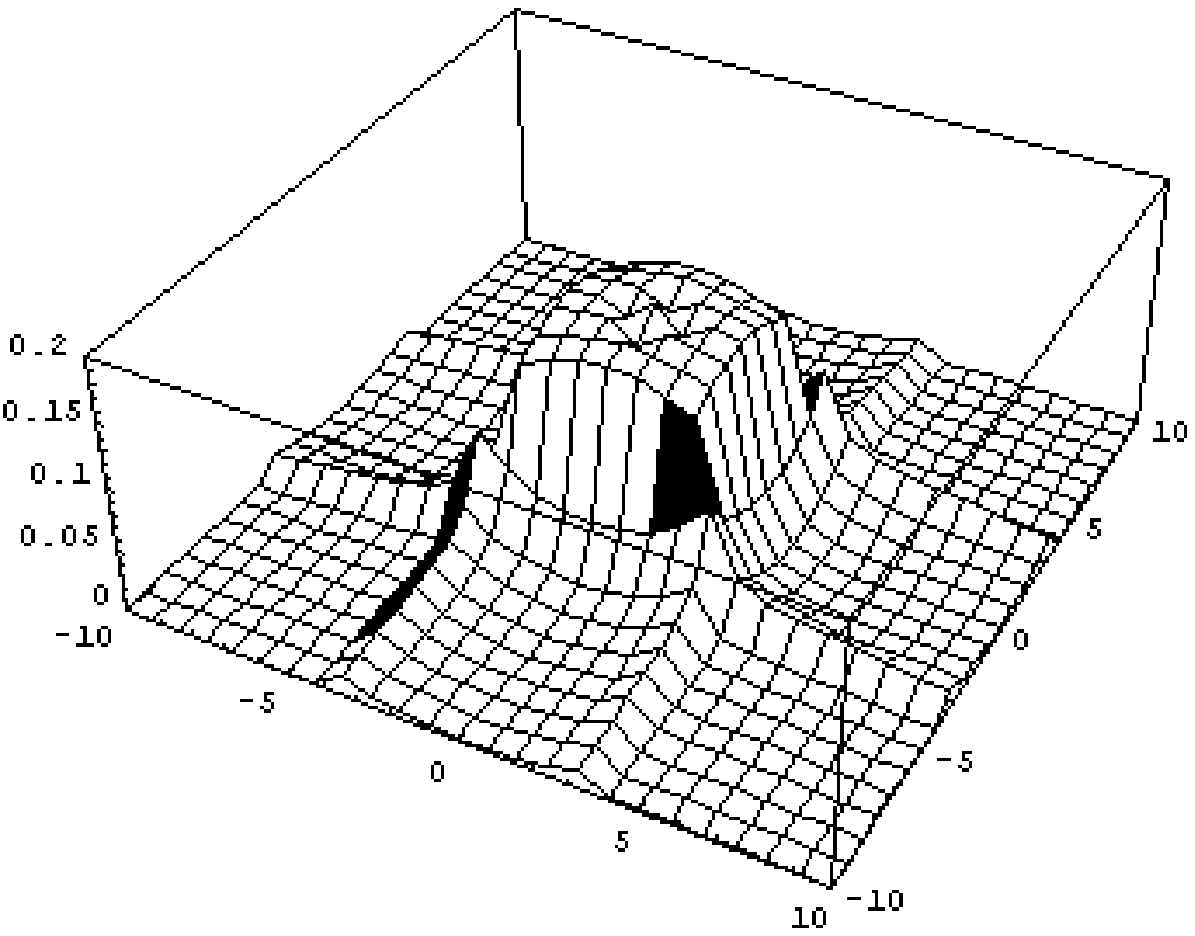}}\hspace{6mm}
\hfill\\
\begin{figure}[h]
\caption{\label{fig2} $|\Psi^{(o)}(\underline x)|$ for
$h_1(x)=h_2(x)=\frac{\sin(\pi x)}{\pi x}$}
\end{figure}
\end{center}

We postpone other details of this construction to the next
subsection, since in that situation the kernel will have a simpler
form and this will simplify all our computations.

\subsection{Another example from quantum mechanics}

The example we discuss now arises from an old paper, \cite{bag2}, in
which some aspects of the relevance of o.n. wavelet bases in the
analysis of the quantum Hall effect were investigated. In particular
we have constructed a two dimensional quantum hamiltonian, close to
that of the quantum Hall effect, which again can be written in new
variables as the hamiltonian of a quantum oscillator. This was
useful to show, via explicit computation of the Coulomb energy of a
two-electrons system, that wavelet-like o.n. bases in the lowest
Landau levels have very good localization features when compared to
bases arising from gaussian functions, \cite{bag2}. In this case the
canonical transformation is \be \hat x_1^{(n)}=\hat p_1^{(o)}+\hat
p_2^{(o)},\quad \hat x_2^{(n)}=\hat p_2^{(o)},\quad \hat
p_1^{(n)}=-\hat x_1^{(o)},\quad \hat p_2^{(n)}=\hat x_1^{(o)}-\hat
x_2^{(o)},\quad \label{43}\en and  we find that, \cite{bag2}, \be
K(\underline x;\underline
y)=\frac{1}{2\pi}\,\exp\left\{i\left(x_1y_1+y_2(x_2-x_1)\right)\right\}\label{44}\en
If we compute again $ \Psi^{(o)}(\underline x)=\int\,d\underline
y\,K(\underline x;\underline y)\,\Psi^{(n)}(\underline y) $ for
$\Psi^{(n)}(\underline y)=h_1(y_1)h_2(y_2)$ we have several
different possibilities, which are considered in the following.

\vspace{2mm}

{\bf Case 1:} $h_1(x)=h_2(x)=\chi_{[0,1]}(x)$.

In this case the computation is  easily performed and we get:
$$\Psi^{(o)}(\underline
x)=\frac{1}{2\pi}\,\frac{e^{ix_1}-1}{x_1}\,\frac{e^{i(x_2-x_1)}-1}{x_2-x_1},$$
which is clearly {\bf not} the product of a function which depends
on $x_1$ and another function which depends on $x_2$. This is
another evidence of the fact that, as we have stated many times
along the text, our approach is not merely a {\em tensor product technique}
but really produces different functions which are genuinely in
$\Lc^2(\Bbb{R}^2)$ and not necessarily in
$\Lc^2(\Bbb{R})\otimes\Lc^2(\Bbb{R})$.

We also want to investigate, in the ambit of this Example, the role
of the operators $T_{\pm}$. We do it here because, in view of the
simplicity of (\ref{43}), it is easier to display the result. Let us
recall that, because of (\ref{31}), the entire set of functions in
$\Lc^2(\Bbb{R}^2)$ looks like $\Psi^{(o)}_{\underline l,\underline
k}(\underline x)=T_+(\underline a(\underline l))\,T_-(\underline
b(\underline k))\,\Psi^{(o)}(\underline x)$, where $T_+(\underline
a(\underline l))=e^{i\underline{\hat X}^{(n)}\cdot\underline
a(\underline l)}=e^{i\hat X_1^{(n)}a_1l_1}\,e^{i\hat
X_2^{(n)}a_2l_2}$ and $T_-(\underline b(\underline
k))=e^{i\underline{\hat P}^{(n)}\cdot\underline b(\underline
k)}=e^{i\hat P_1^{(n)}b_1k_1}\,e^{i\hat P_2^{(n)}b_2k_2}$. Moreover,
because of the transformation (\ref{43}) (and of its counterpart for
the capital operators), it is a standard computation to check that
\be \Psi^{(o)}_{\underline l,\underline k}(\underline
x)=e^{i(x_1+a_1l_1)(b_2k_2-b_1k_1)}\,e^{-i(x_2+\underline
a\cdot\underline l)b_2k_2}\Psi^{(o)}(\underline x+\underline
a\cdot\underline l) \label{45bis}\en which is again the modulated
and translated version of $\Psi^{(o)}(\underline x)$, but in a
slightly non-trivial way.

\vspace{2mm}

{\bf Case 2:} $h_1(x)=h_2(x)=\frac{\sin(\pi x)}{\pi x}$.

This is again a symmetrical choice for which all the computations
can be easily performed. The result can be written as
$$
\Psi^{(o)}(\underline
x)=\frac{1}{2\pi}\,\chi_{[-\pi,\pi]}(x_1)\,\chi_{[-\pi,\pi]}(x_2-x_1)
$$
which is again not-factorizable. The next figure displays the plot
of this function, which is clearly compactly supported. As for the
functions $\Psi^{(o)}_{\underline l,\underline k}(\underline x)$,
since they are related to the same canonical transformation, then
they can be obtained as in Case 1 and look as in (\ref{45bis}).

\begin{center}
\mbox{\includegraphics[height=6.4cm, width=9.0cm]
{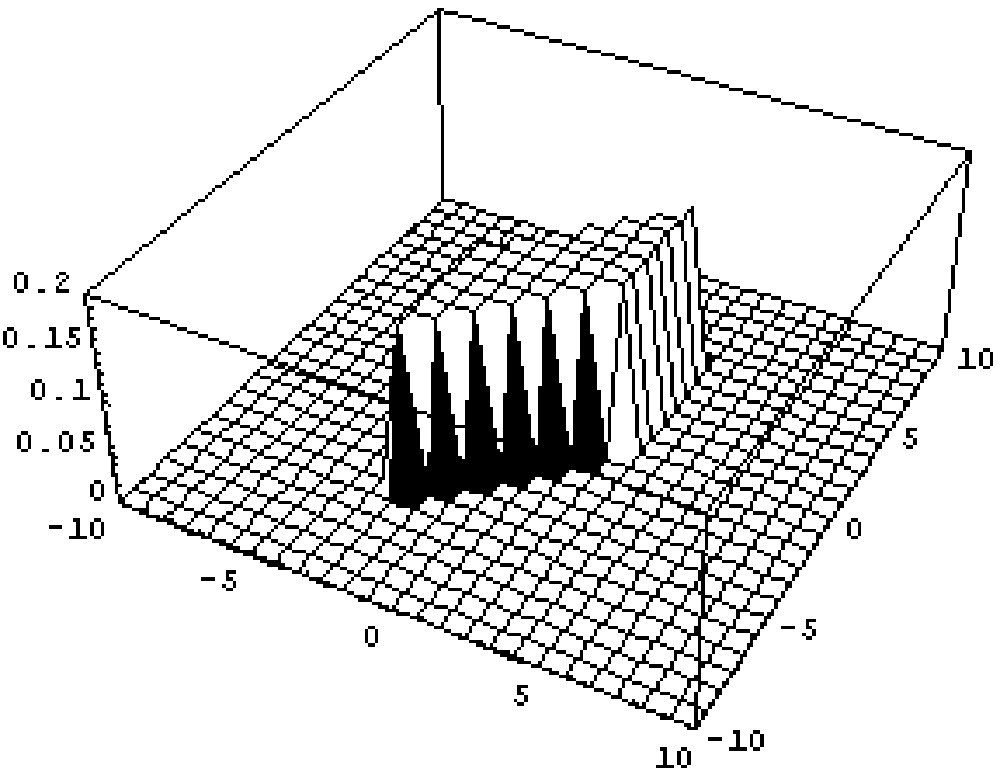}}\hspace{6mm}
\hfill\\
\begin{figure}[h]
\caption{\label{fig3} $|\Psi^{(o)}(\underline x)|$ for
$h_1(x)=h_2(x)=\frac{\sin(\pi x)}{\pi x}$ }
\end{figure}
\end{center}

\vspace{2mm}

{\bf Case 3:} $h_1(x)=\frac{\sin(\pi x)}{\pi x}$ and
$h_2(x)=\chi_{[0,1]}(x)$.

This choice is no longer symmetric and the result, which again can
be easily computed, is the following:
$$
\Psi^{(o)}(\underline
x)=\frac{1}{2\pi}\,\chi_{[-\pi,\pi]}(x_1)\,\frac{e^{i(x_2-x_1)}-1}{x_2-x_1}.
$$
Once again, the functions $\Psi^{(o)}_{\underline l,\underline
k}(\underline x)$ can be computed as (\ref{45bis}).

\subsection{An example in $d=3$}

As we have widely discussed before the only real ingredient to apply
our procedure is a canonical transformation from three {\em old}
pairs of conjugate operators into three {\em new} pairs of similar
operators. No physical motivation is strictly required, and indeed
there is no motivation in the map we consider below, which is chosen
only in view of its simplicity. We put \bea \left\{
    \begin{array}{ll}
\hat x_1^{(n)}=\frac{1}{2}\left(\hat p_3^{(o)}-\hat p_1^{(o)}-\hat p_2^{(o)}\right),\qquad
\hat p_1^{(n)}=\hat x_1^{(o)}+\hat x_2^{(o)}\\
\hat x_2^{(n)}=\frac{1}{2}\left(\hat p_1^{(o)}-\hat p_2^{(o)}-\hat
p_3^{(o)}\right),\qquad
\hat p_2^{(n)}=\hat x_2^{(o)}+\hat x_3^{(o)}\\
\hat x_3^{(n)}=\frac{1}{2}\left(\hat p_2^{(o)}-\hat p_1^{(o)}-\hat
p_3^{(o)}\right),\qquad
\hat p_3^{(n)}=\hat x_3^{(o)}+\hat x_1^{(o)}\\
       \end{array}
        \right.
        \label{45}
 \ena
In this case the kernel is, \cite{mo}, \be K(\underline x;\underline
y)=\frac{1}{2\pi^{3/2}}\,\exp\left\{-i\left(y_1(x_1+x_2)+y_2(x_2+x_3)+y_3(x_1+x_3)
\right)\right\}\label{46}\en

If we now  choose  $\Psi^{(n)}(\underline
y)=\chi_{[0,1]}(y_1)\,\chi_{[0,1]}(y_2)\,\chi_{[0,1]}(y_3)$ we get
$$
\Psi^{(o)}(\underline
x)=\frac{i}{2\pi^{3/2}}\,\frac{e^{-i(x_1+x_2)}-1}{x_1+x_2}\,
\frac{e^{-i(x_2+x_3)}-1}{x_2+x_3}\,\frac{e^{-i(x_1+x_3)}-1}{x_1+x_3}
$$
Also this function is not the product of
three one-dimensional functions, as claimed before. From this
$\Psi^{(o)}(\underline x)$ we can further deduce the form of the
functions $\Psi^{(o)}_{\underline l,\,\underline k}(\underline x)$,
following the same steps as in Section IV.2. We leave to the reader
the construction of other examples which can be obtained, for
instance, considering different choices of the single functions
defining $\Psi^{(n)}(\underline y)$.

\subsection{Two frames in $d=2$}

Our ingredient for this example is the following: $h(t)\equiv
h_1(t)=h_2(t)=\frac{e^{-t^2/2}}{\pi^{1/4}}$, which is such that the
set $e^{im\omega_o t}h(t-nt_o)$ is a frame for all possible choices
$\omega_ot_o\leq 2\pi$.

In this case, if we compute $\Psi^{(o)}(\underline x)$ using the
kernel (\ref{42}), we get $$\Psi^{(o)}(\underline
x)=\frac{1}{\sqrt{2\pi}}\,e^{-(x_1^2+x_2^2)/4}$$ This result is not
unexpected since it was well known since many years, see \cite{bms}
and references therein, the role of this two-dimensional gaussian in
the analysis of the Landau levels arising in the analysis of the
quantum Hall effect.

If, on the other way, we choose the kernel in (\ref{44}) we find,
after few easy computations, $$\Psi^{(o)}(\underline
x)=\frac{1}{\sqrt{2\pi}}\,e^{-(x_1^2+(x_2-x_1)^2)/2}$$ which shows
an asymmetric behavior between the variables $x_1$ and $x_2$.

\section{What about wavelet?}

The procedure discussed so far for Gabor-like systems can be
extended to sets of wavelets. To deduce this extension it is
convenient to give first few mathematical results, complementing the
ones discussed in Section II, and this will be done in the next
subsection, while the explicit extension with two examples will be
discussed in Section V.2.

\subsection{Still more mathematics}

In the case of wavelets the modulation operator in Section II must
be replaced by the dilation operator while the translation operator
is still needed. As before, in our approach is convenient to pay
attention to the difference between the Hilbert spaces $\Hil$ and
$\Lc^2(\Bbb{R}^2)$. In this case it is maybe more natural to start
defining the operators on $\Lc^2(\Bbb{R}^2)$ and then defining their
{\em small counterparts} acting on $\Hil$. Notice that this is
exactly the opposite of what we have done in Section II. For this we
begin defining the following operators, \cite{chr}: \bea \left\{
    \begin{array}{ll}
        T_1f^{(n)}(x_1,x_2)=f^{(n)}(x_1-1,x_2), \\
        T_2f^{(n)}(x_1,x_2)=f^{(n)}(x_1,x_2-1), \\
        D_1f^{(n)}(x_1,x_2)=\sqrt{2}\,f^{(n)}(2x_1,x_2), \\
        D_2f^{(n)}(x_1,x_2)=\sqrt{2}\,f^{(n)}(x_1,2x_2), \\
       \end{array}
        \right.\qquad \forall f^{(n)}(x_1,x_2)=<\xi_{\underline x}^{(n)},f>_{\Hil}\in\Lc^2(\Bbb{R}^2)
        \label{51}
 \ena

Once again we only need to introduce operators which act as
dilations and translations in the {\em new} variables. This is
exactly the analogous of what we have done in Section II.

It is trivial to check that all these operators are unitary, that
$[T_1,T_2]=[D_1,D_2]=[T_1,D_2]=[T_2,D_1]=0$, and that \bea \left\{
    \begin{array}{ll}
T_m^k\,D_m^j=D_m^j\,T_m^{2^jk}\\
D_m^j\,T_m^{k}=T_m^{2^{-j}k}\,D_m^j\\
       \end{array}
        \right.\qquad m=1,2,\quad\forall j,\,k\in\Bbb{Z}
        \label{52}
 \ena
If we now introduce the {\em small operators} $t_m$ and $d_m$
in the same spirit and repeating the same comments as  in (\ref{210}),
 \be
 T_m\,<f,g>_{\Hil}=:<f,\,t_mg>_{\Hil}, \qquad
D_m\,<f,g>_{\Hil}:=<f,\,d_mg>_{\Hil},\label{53}\en for  $m=1,2$,  it is easy to
deduce that \bea \left\{
    \begin{array}{ll}
t_1\xi_{\underline x}^{(n)}=\xi_{(x_1+1,x_2)}^{(n)},\qquad\,\,\,
t_2\xi_{\underline x}^{(n)}=\xi_{(x_1,x_2+1)}^{(n)}\\
d_1\xi_{\underline
x}^{(n)}=\frac{1}{\sqrt{2}}\,\xi_{\left(\frac{x_1}{2},x_2\right)}^{(n)},\qquad
d_2\xi_{\underline x}^{(n)}=\frac{1}{\sqrt{2}}\,\xi_{\left(x_1,\frac{x_2}{2}\right)}^{(n)}\\
       \end{array}
        \right.
        \label{54}
 \ena
which is all that we need in the following.

\subsection{Wavelets o.n. bases in any dimension}

The problem we want to solve here is completely analogous to the one
considered in Section III. For simplicity we fix here $d=2$, but the
extension to higher dimensions is straightforward. More in details
we want to construct a square-integrable function, which we again
call $\Psi^{(o)}(\underline x)$, such that the set \be
\F=\left\{\Psi^{(o)}_{\underline j,\underline k}(\underline
x)=D_1^{j_1}\,D_2^{j_2}\,T_1^{k_1}\,T_2^{k_2}\,\Psi^{(o)}(\underline
x),\,\underline k,\underline j\in\Bbb{Z}^2\right\} \label{55}\en is
an o.n. basis in $\Lc^2(\Bbb{R}^2)$.

Before starting with our construction, it is worth noticing that as
before this set is only apparently the set of dilated and translated
of what we still call the {\em mother wavelet}
$\Psi^{(o)}(\underline x)$. In other words, even if the operators
$D_j$ and $T_j$ look like dilation and translation operators, they
behave as these operators only in the {\em new} variables. Hence, in
general, we have $\Psi^{(o)}_{\underline j,\underline k}(\underline
x)\neq 2^{(j_1+j_2)/2}\,\Psi^{(o)}(2^{j_1}
x_1-k_1,\,2^{j_2}x_2-k_2)$ (but for trivial canonical
transformations, of course). Nevertheless, it is not hard to compute
the way in which the operators $D_j$ and $T_j$ act on
$\Psi^{(o)}(\underline x)$. The strategy adopted is completely
identical to the one in (\ref{32}) and produces,  making use of
(\ref{54}), \be \Psi^{(o)}_{\underline j,\underline k}(\underline
x)=2^{(j_1+j_2)/2}\,\int_{\Bbb{R}^2}\,d\underline y\,K(\underline
x;\underline y)\,\Psi^{(n)}(2^{\underline j}\,\underline
y-\underline k), \label{56}\en where $\underline j,\,\underline
k\in\Bbb{Z}^2$ and we have introduced the compact notation
$2^{\underline j}\,\underline y-\underline k=(2^{j_1}
x_1-k_1,\,2^{j_2}x_2-k_2)$.

First of all we want these functions to be mutually o.n. Once again
we start considering the following scalar product in
$\Lc^2(\Bbb{R}^2)$: $$\tilde\Omega_{\underline j,\,\underline
k}:=<\Psi^{(o)}_{\underline j,\underline
k},\Psi^{(o)}>_{\Lc^2(\Bbb{R}^2)}=\int_{\Bbb{R}^2}\,d\underline
x\,\overline{\Psi^{(o)}_{\underline j,\underline k}(\underline
x)}\,\Psi^{(o)}(\underline x)$$ which, using (\ref{56}) and the
resolution of the identity for $\xi_{\underline y}^{(n)}$, can be
written as
$$
\tilde\Omega_{\underline j,\,\underline
k}=2^{(j_1+j_2)/2}\,\int_{\Bbb{R}^2}\,d\underline
y\,\overline{\Psi^{(n)}(2^{\underline j}\,\underline y-\underline
k)}\,\Psi^{(n)}(\underline y)
$$
Furthermore, if we start with a factorizable $\Psi^{(n)}(\underline
y)$, $\Psi^{(n)}(\underline y)=h_1(y_1)\,h_2(y_2)$, this can be
rewritten as \be \tilde\Omega_{\underline j,\,\underline
k}=\left(2^{j_1/2}\int_{\Bbb{R}}dy_1\,\overline{h_1(2^{j_1}y_1-k_1)}\,h_1(y_1)\right)
\left(2^{j_2/2}\int_{\Bbb{R}}dy_2\,\overline{h_2(2^{j_2}y_2-k_2)}\,h_2(y_2)\right)
\label{57}\en In analogy with what we have done for Gabor systems,
we now assume that the two sets of functions
$\E_l=\{2^{j/2}h_l(2^jx-k),\,j,k\in\Bbb{Z}\}$, $l=1,2$, are both
o.n. wavelets bases of $\Lc^2(\Bbb{R})$. If this is so, then it is
clear that $\tilde\Omega_{\underline j,\,\underline
k}=\delta_{\underline j,\,\underline 0}\,\delta_{\underline
k,\,\underline 0}$. This, in turns, because of the commutation rules
in (\ref{52}),  implies that \be <\Psi^{(o)}_{\underline
j,\underline k},\Psi^{(o)}_{\underline n,\underline
m}>_{\Lc^2(\Bbb{R}^2)}=\delta_{\underline j,\,\underline
n}\,\delta_{\underline k,\,\underline m}\label{58}\en which is what
we had to prove first.

The next step is to prove the completeness in $\Lc^2(\Bbb{R}^2)$ of
the set $\F$. This is guaranteed by the assumption that both $\E_1$
and $\E_2$ are complete in $\Lc^2(\Bbb{R})$, and the proof is left
to the reader since it does not differ significantly from the one
given for Gabor systems. Therefore $\F$ is an o.n. wavelet-like
basis in $\Lc^2(\Bbb{R}^2)$.

As already mentioned the extension of the procedure to $d>2$ is not
particularly difficult and follows the same steps as for the
Gabor-like basis.

\vspace{2mm}

{\bf Example:} $h_1(x)=h_2(x)=H(x)$, where $H(x)$ is the Haar mother
wavelet.

We consider first the choice (\ref{42}) of the kernel, which
corresponds to the canonical transformation arising from the quantum
Hall effect. In this case we get
$$
\Psi^{(o)}(\underline
x)=\frac{i\,e^{-ix_1x_2/2}}{2\pi}\,\left(\int_0^{1/2}\,e^{ix_1y_1}\,\frac{\left(e^{i(x_2-y_1)}-1\right)^2}
{x_2-y_1}\,dy_1-\int_{1/2}^{1}\,e^{ix_1y_1}\,\frac{\left(e^{i(x_2-y_1)}-1\right)^2}
{x_2-y_1}\,dy_1\right)
$$
and Figure 3 shows the plot of its modulus. In this case, finding
the action of the operators $D_j$ and $T_j$ on
$\Psi^{(o)}(\underline x)$ is harder than in Section III, and
formula (\ref{56}) should be used.

\begin{center}
\mbox{\includegraphics[height=6.4cm, width=9.0cm]
{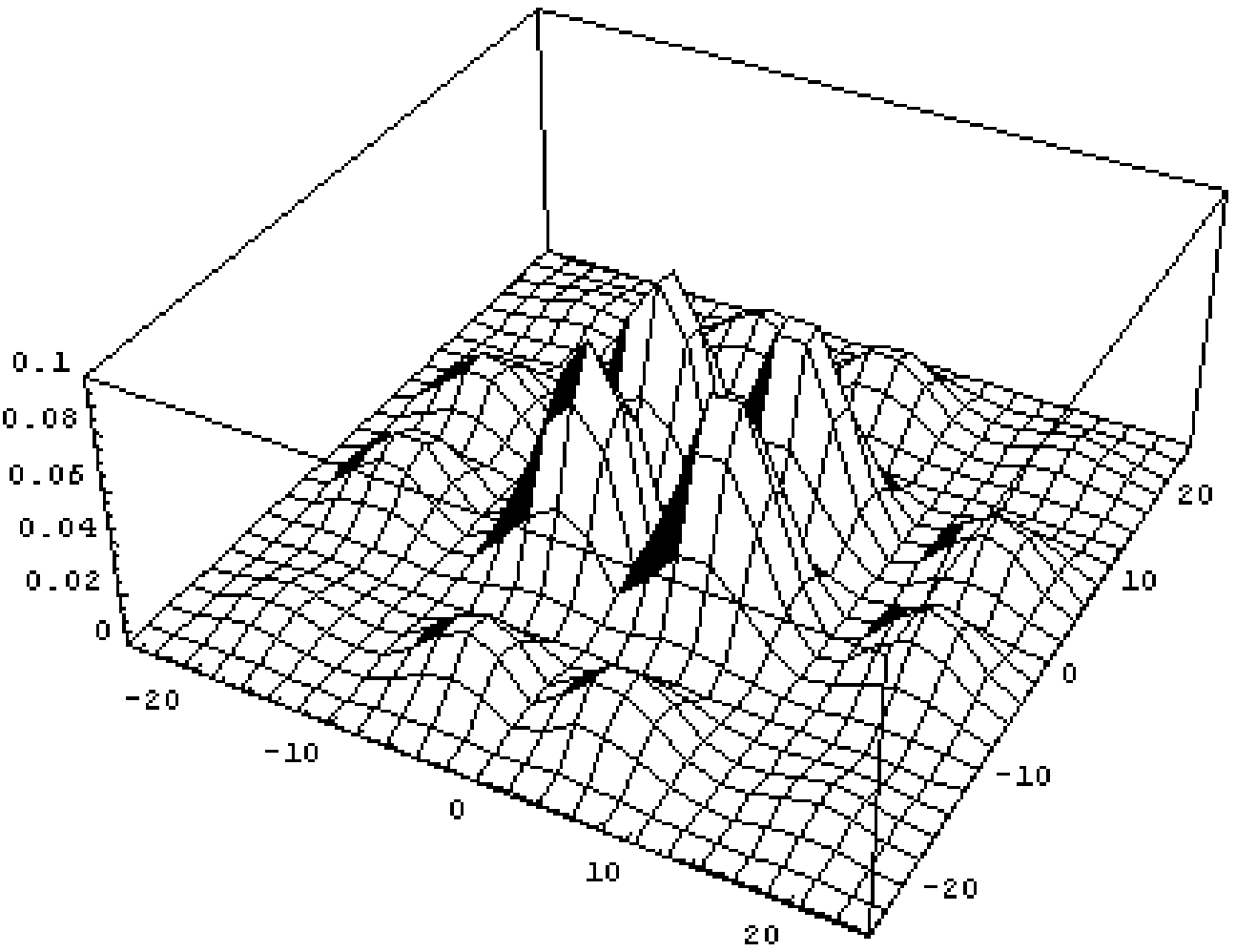}}\hspace{6mm}
\hfill\\
\begin{figure}[h]
\caption{\label{fig4} $|\Psi^{(o)}(\underline x)|$ for
$h_1(x)=h_2(x)=H(x)$ }
\end{figure}
\end{center}

Much easier is the computation of $\Psi^{(o)}(\underline x)$ using
the other two-dimensional kernel, (\ref{44}). In this case we find
$$
\Psi^{(o)}(\underline
x)=\frac{1}{2\pi\,x_1(x_1-x_2)}\,\left(e^{ix_1/2}-1\right)^2\left(e^{i(x_2-x_1)/2}-1\right)^2
$$
which, as expected, is not factorized. Again,
$\Psi^{(o)}_{\underline j,\underline k}(\underline x)$ is given by
(\ref{56}).

\section{Conclusions}

In this paper we have constructed an easy technique, based on
quantum canonical transformations, which allow us to construct a
function  $\Psi^{(o)}(\underline x)$ in $\Lc^2(\Bbb{R}^d)$ out of
$d$ functions in $\Lc^2(\Bbb{R})$ such that:

\begin{itemize}

\item if these functions generate  Gabor o.n. bases in
$\Lc^2(\Bbb{R})$, then $\Psi^{(o)}(\underline x)$ generates a
Gabor-like o.n. basis in $\Lc^2(\Bbb{R}^d)$;

\item if these functions generate only  Gabor frames in
$\Lc^2(\Bbb{R})$, then $\Psi^{(o)}(\underline x)$ generates a
Gabor-like frame in $\Lc^2(\Bbb{R}^d)$;

\item if these functions generate  wavelet o.n. bases in
$\Lc^2(\Bbb{R})$ then $\Psi^{(o)}(\underline x)$ generates a
wavelet-like o.n. basis in $\Lc^2(\Bbb{R}^d)$.

\end{itemize}

The reason why we call these sets {\em Gabor-like} and {\em
wavelet-like} rather than simply {\em Gabor} and {\em wavelet} is
related to the fact that we are using, as we have explained above,
translation, modulation and dilation operators in the {\em new}
variables rather than in the {\em old} ones. These  are still
unitary operators but act on $\Psi^{(o)}(\underline x)$ in a
slightly modified way, and therefore they do not produce the
standard sets of functions.

The next step of our analysis will be concerned with possible
physical applications of our construction, in the attempt of
mimicking and extending our old results on quantum Hall effect.

\section*{Acknowledgements}

This work was  supported by M.U.R.S.T..

\end{document}